\documentstyle[preprint,aps,pra,epsf]{revtex}
\tighten
\draft
\begin{document}
\title{Magnetic structures of RbCuCl$_3$ in a transverse field} 

\author{A. E. Jacobs and T. Nikuni}
\address{Department of Physics, University of Toronto, Toronto, Ontario,
Canada M5S 1A7}

\date{\today}
\maketitle
\begin{abstract}
A recent high-field magnetization experiment found a phase transition of 
unknown character in the layered, frustrated antiferromagnet RbCuCl$_3$, 
in a transverse field (in the layers). 
Motivated by these results, we have examined the magnetic structures 
predicted by a model of RbCuCl$_3$, using the classical approximation. 
At small fields, we obtain the structure already known to be optimal, 
an incommensurate (IC) spiral with wave vector {\bf q} in the layers. 
At higher fields, we find a staircase of long-period commensurate (C) phases 
(separated initially by the low-field IC phase), then two narrow IC phases, 
then a fourth IC phase (also with intermediate C phases), and finally the 
ferromagnetically aligned phase at the saturation field $H_S$. 
The three-sublattice C states familiar from the theory of the triangular 
antiferromagnet are never optimal. 
The C phases and the two intermediate IC phases were previously unknown in 
this context. 
The magnetization is discontinuous at a field $\approx0.4H_S$, in qualitative 
agreement with experiment, though we find much fine structure not reported. 

\end{abstract}
\pacs{75.10.Hk, 75.25.+z, 75.50.Ee}

\section{Introduction}
The ABX$_3$ family of layered compounds has been studied extensively for 
several decades\cite{collins}.  
Much of this interest arises because the frustrated antiferromagnetic 
interactions in the $a$-$b$ layers give rise to unusual magnetic properties. 
Quantum and thermal fluctuations can have exceptionally large effects in 
these materials; 
for example, the stacked triangular antiferromagnet (TAFM) CsCuCl$_3$ 
exhibits novel fluctuation-induced magnetic phase transitions in both a 
longitudinal field\cite{nikuni,motokawa} (in the $c$ direction, normal to the 
layers) and a transverse field \cite{werner,schotte,jacobs}. 

\begin{figure}
\epsfxsize=110mm
\centerline{\epsfbox{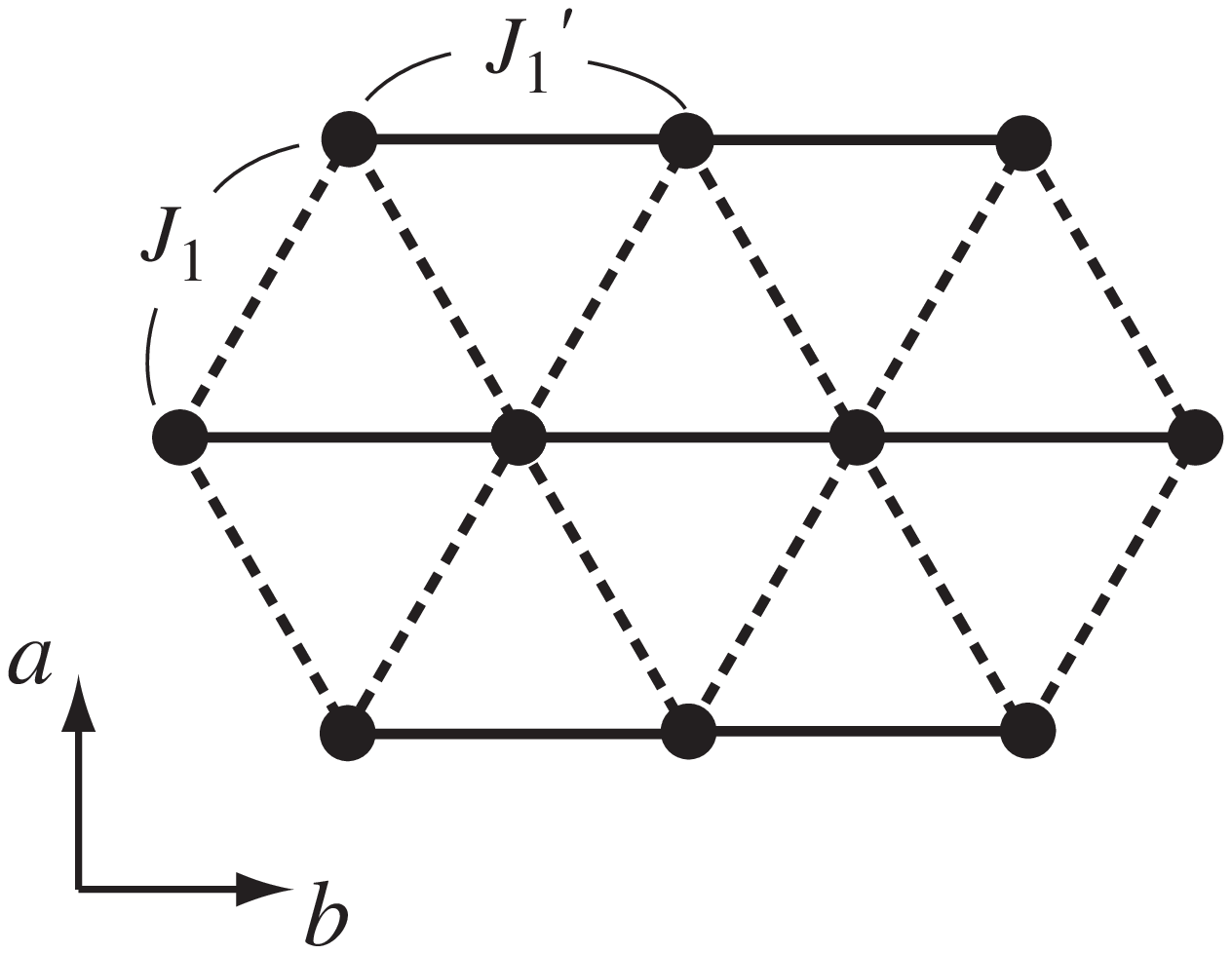}}
\begin{caption}
{The exchange interactions $J_1$ (broken lines) and $J_1'$ (solid lines)
in the distorted triangular lattice in the $a$-$b$ plane.}
\label{lattice}
\end{caption}
\end{figure}

RbCuCl$_3$, another frustrated antiferromagnet of the ABX$_3$ family, is 
magnetically ordered for temperatures $T$ less than $T_N \approx 19$K 
\cite{tazuke}; 
the saturation field $H_S$ (above which the ferromagnetically aligned phase 
is stable) is inconveniently large however ($\approx 66$T) \cite{tanaka}. 
Like CsCuCl$_3$, RbCuCl$_3$ is ferromagnetically stacked (in the $c$ 
direction) and it has an incommensurate (IC) structure in zero field
\cite{reehuis}.
The Cu$^{2+}$ ions in the $a$-$b$ planes of RbCuCl$_3$ do not however form 
a regular triangular lattice. 
A room-temperature structural phase transition\cite{crama,harada} distorts 
the structure, yielding a spatially anisotropic intraplane exchange 
(see Figure~\ref{lattice}). 
As a result, at zero field the spins do not adopt the familiar 
three-sublattice, $120^\circ$ structure of the TAFM; 
the structure is instead an IC helical structure along the $b$ direction
with rotation angle $108^{\circ}$ between adjacent spins\cite{Mn}. 

A neutron-diffraction experiment\cite{reehuis} at $T=1.6$K in zero field
found that the spins lie in the $a$-$b$ planes and are ferromagnetically 
aligned in the $c$ direction. 
The magnetic properties in large external field were recently investigated 
by means of magnetization measurements and by ESR\cite{tanaka}. 
A discontinuity was found in the magnetization at a transverse field of 
$H=21.2$T $(\approx 0.32H_S)$. 
The nature of this field-induced magnetic transition and the structure of the 
high-field state could however not be determined in these experiments. 

The following reports theoretical results for spin structures of RbCuCl$_3$ 
in a transverse magnetic field. 
We use the simplest model Hamiltonian and investigate it analytically and 
numerically within the classical approximation. 
In the field range below $H_S$, solution of the Euler-Lagrange equations 
yields four IC phases, and in addition many long-period commensurate 
(C) phases in both the low-field and high-field regions; 
the three-sublattice C states of the TAFM are never optimal. 
Of these C and IC phases, only the low-field IC (helical) phase and
the high-field IC (fan) phase were known previously \cite{nagamiya}; 
the two intermediate IC phases occupy however only a very small field range. 
The magnetization is discontinuous, in qualitative agreement with experiment; 
the discontinuity occurs however at $H=0.41H_S$, rather than at $0.32H_S$ as 
observed, and the magnetization shows additional fine structure. 

\section{Model Hamiltonian}

The model Hamiltonian describing RbCuCl$_3$ in a transverse magnetic 
field (which we take to lie in the $x$ direction, to be explicit) is 
\cite{tanaka} 
\begin{equation}
{\cal H}=-\sum_{in}2J_0{\bf S}_{in}\cdot {\bf S}_{in+1}
-\sum_{in}2\Delta J_0 S^z_{in}S^z_{in+1}
-\sum_{\langle ij\rangle,n}2J_1^{ij}{\bf S}_{in}\cdot{\bf S}_{jn}
-g\mu_{\rm B}H\sum_{in}S^x_{in} .
\label{eq:H}
\end{equation}
The $x$ and $z$ axes are taken along the $a$ and $c$ directions respectively; 
${\bf S}_{in}$ is the spin operator ($S={1\over2})$ of the $i$th Cu$^{2+}$ 
site in the $n$th $a$-$b$ plane, and the $\langle ij \rangle$ sum runs over 
nearest-neighbor sites in the $a$-$b$ plane. 
The first term in Eq.~(\ref{eq:H}) is the nearest-neighbor ferromagnetic 
exchange interaction ($J_0>0$) in the $c$ direction; 
the second, with $\Delta J_0 <0$, describes an anisotropy of easy-plane type. 
The third term, with $J_1^{ij}<0$, is the antiferromagnetic exchange 
interaction in the $a$-$b$ planes. 
As shown in Figure~\ref{lattice}, $J_1^{ij}$ is anisotropic; 
$J_1^{ij}=-J'_1$ along the $b$ direction while
$J_1^{ij}=-J_1$ in the other two directions.
The fourth and last term is the Zeeman energy. 
The parameters in Eq.~(\ref{eq:H}) are
$J_0=25.7$K, $\Delta J_0=-0.45$K, $J_1=10.6$K, $J_1'=17.4$K, and 
$g=2.14$ (for $H\perp c$). 

We investigate this Hamiltonian in the classical approximation. 
The spin operators ${\bf S}_{in}$ then become classical vectors of length 
${1\over2}$, and the Hamiltonian becomes an energy function; 
the ground-state spin structures are those that minimize the energy. 
The classical ground state of Eq.~(\ref{eq:H}) in zero field is the
helical structure~\cite{nagamiya}: 
\begin{equation}
{\bf S}_{in}=S\hat{\bf x}\cos({\bf q}\cdot{\bf r}_{in})+
S\hat{\bf y}\sin({\bf q}\cdot{\bf r}_{in});
\end{equation}
the easy-plane anisotropy favors spins in the $a$-$b$ planes. 
The optimal wave vector is ${\bf q}=(0,q_0,0)$, with $\cos q_0=-J_1/2J_1'$; 
the above parameter values give $q_0/2\pi=0.2993$. 

In the special case $J_1'=J_1$, Eq.~(\ref{eq:H}) reduces to the Hamiltonian 
of the stacked TAFM. 
Classically, the TAFM ground states are three-sublattice structures; 
they evolve continuously with field $H$, from the zero-field $120^\circ$ 
structure (with $q_0/2\pi=1/3$ in the above picture) up to the aligned state 
at $H_S$. 
These states are continuously degenerate at all $H<H_S$; 
energy minimization gives only two relations for the three angles $\phi_i$ 
between the sublattice spins and the field, namely 
\begin{equation}
\cos\phi_1+\cos\phi_2+\cos\phi_3= g\mu_{\rm B}H/(6 J_1 S)
\label{commen}
\end{equation}
and $\sum_{i=1}^3 \sin \phi_i=0$. 
This degeneracy, which is nontrivial, is broken identically by thermal and 
quantum fluctuations omitted in the classical approximation
\cite{kawamura,chubukov}. 

We determine the spin structures of RbCuCl$_3$ in transverse fields with the 
help of two reasonable assumptions: the spins remain in the $a$-$b$ planes 
and they are aligned ferromagnetically along the $c$ and $a$ directions.
We apply periodic boundary conditions in the $b$ direction, to reduce 
finite-size effects. 
Then, from Figure 1, we can use a chain description in which the classical 
energy per spin is 
\begin{equation}
{\cal E}(\{\phi_l\})=\frac{1}{L}\sum_{l=1}^L
\left[ 4 J_1 S^2\cos(\phi_l-\phi_{l+1}) 
      +  2 J_1'S^2\cos(\phi_l-\phi_{l+2})
      -g\mu_{\rm B}HS \cos\phi_l \right];
\label{chain}
\end{equation}
the chain-site index $l$ runs in the $b$ direction, $\phi_l$ is the angle 
between ${\bf S}_l$ and the field, and the chain length $L$ is a multiple of 3. 
Equation (\ref{chain}) is just the classical energy per spin for the axial 
next-nearest-neighbour Heisenberg chain, except for the restriction on $L$. 
The optimal solutions are never TAFM states, but they have related features, 
and so we use also the representation $\phi_{i,l'}\equiv\phi_{3l'+i}$ where 
$i=1,2,3$ is the sublattice index.

The Euler-Lagrange equations are obtained from 
$\partial {\cal E}/\partial \phi_l=0$: 
\begin{eqnarray}
  4 J_1  S^2\left[ \sin(\phi_l    -\phi_{l+1}) 
                  -\sin(\phi_{l-1}-\phi_l) \right]
   &+2 J_1' S^2\left[ \sin(\phi_l    -\phi_{l+2}) 
                  -\sin(\phi_{l-2}-\phi_l) \right]\nonumber \\
  &=g \mu_B H S \sin\phi_l \ .
\label{Euler}
\end{eqnarray}
These are $L$ coupled, nonlinear difference equations with boundary 
conditions $\phi_{l+L}=\phi_l\pmod{2\pi}$. 
It is not in general legitimate to approximate them by differential 
equations; 
doing so would miss the sequences of C states that we find at both 
low and high fields. 

Equations (\ref{Euler}) have several analytical solutions, namely the 
zero-field helical state $\phi_l=ql$, the aligned state $\phi_l\equiv 0$, 
and the TAFM states described by Eq.~(\ref{commen}) with $J_1$ replaced by 
$\bar J\equiv (2J_1+J'_1)/3$. 
Analytical progress is possible for small fields and for $H$ just below 
$H_S$ \cite{nagamiya}. 
In general Eq.~(\ref{Euler}) must be solved numerically; 
this led us to previously unknown solutions. 
The equations have in fact a multitude of solutions. 
This multiplicity results in part from a degeneracy not present in systems 
(like CsCuCl$_3$) where the IC phase is driven by a Dzyalshinskii-Moriya 
interaction. 
In RbCuCl$_3$, both right- and left-handed helices are degenerate in energy, 
at nonzero field as well as at $H=0$ (where the relation 
$\cos q_0=-J_1/2J_1'$ determines only the magnitude of $q_0$).  
The multiplicity results also from the many possible periods for fixed 
winding number (defined below). 

We solved the equations by Newton's method, starting from approximations 
obtained by various means (for example solutions at a nearby field), 
using periods $L$ up to $\approx 3\times10^4$; 
we linearized the equations about the starting values and solved the linear 
equations to obtain the corrections, repeating the procedure until the 
root-mean-square residual was less than $10^{-11}$ or so. 
The field range from 0 to $H_S$ was covered in increments of 
$\Delta H\approx 2\times 10^{-4}H_S$. 
Properties of the IC states (such as the optimal chain length, the energy, 
and the magnetization) are easily obtained by interpolation; 
for this purpose it is useful to work with lengths corresponding to 
many periods. 
Periodic boundary conditions necessarily give structures with rational 
periods, but inspection usually suffices to distinguish commensurate from 
incommensurate structures. 

\section{Magnetic structures}
The effect of a transverse magnetic field on helical spin structures arising
from competing exchange interactions, such as described by the classical
Hamiltonian in Eq.~(\ref{chain}), was studied analytically in detail by 
Nagamiya {\it et al}. \cite{nagamiya}. 
They found a distorted helical phase at low fields and an IC fan phase at 
high fields. 
In the latter, the spins oscillate about the field direction; 
the oscillations decrease with increasing field, vanishing continuously 
at $H_S$.
If the transition from the low-field helical phase occurs directly to the 
high-field fan phase, then it is 
first-order and occurs at a reduced field estimated as \cite{nagamiya} 
$H_c/H_S=\sqrt{(1+\beta)(2+\beta)}-(1+\beta)$ with $\beta=(1-J_1/J_1')^2$, 
or $H_c=0.423H_S$ for RbCuCl$_3$. 
The structures in intermediate fields depend however on the ratio $J_1/J'_1$.
For $J_1/J'_1\lesssim 1$, as in RbCuCl$_3$, a TAFM state has lower energy 
than either IC phase for fields near ${1\over2}H_S$. 

Our numerical solution of the Euler-Lagrange equations (\ref{Euler}) gives 
the helical IC and fan IC phases at small and large fields respectively, 
as in Ref.~\cite{nagamiya}. 
Our results differ however at intermediate fields, from $H=0.29H_S$ to 
$0.64H_S$. 
For $0.29<H/H_S<0.41$, we find many helical commensurate (C) phases not 
previously known in this context. 
These are not TAFM states (with period 3 in the chain description) but rather 
lock-in phases with larger periods; 
they alternate with the helical IC phase, and within our field resolution 
($\Delta H \approx 2\times 10^{-4} H_S$) eventually supplant it completely. 
The behavior is similar in some respects to that found in the axial, 
next-nearest-neighbour Ising model \cite{bak}. 
At $H=0.41H_S$, a strong first-order transition occurs to a previously 
unknown, second IC phase (we call it the IC2 phase); 
this appears to be the transition observed in Ref. \cite{tanaka}. 
Two weaker first-order transitions follow swiftly, to a third IC phase (the 
IC3 phase, also previously unknown) and then to fan phases (both C and IC, 
intermingled). 
Finally, there is a second-order transition to the aligned phase; 
we derive an analytical expression for the saturation field, in terms of the 
parameters in the Hamiltonian. 

\begin{figure}
\epsfxsize=110mm
\centerline{\epsfbox{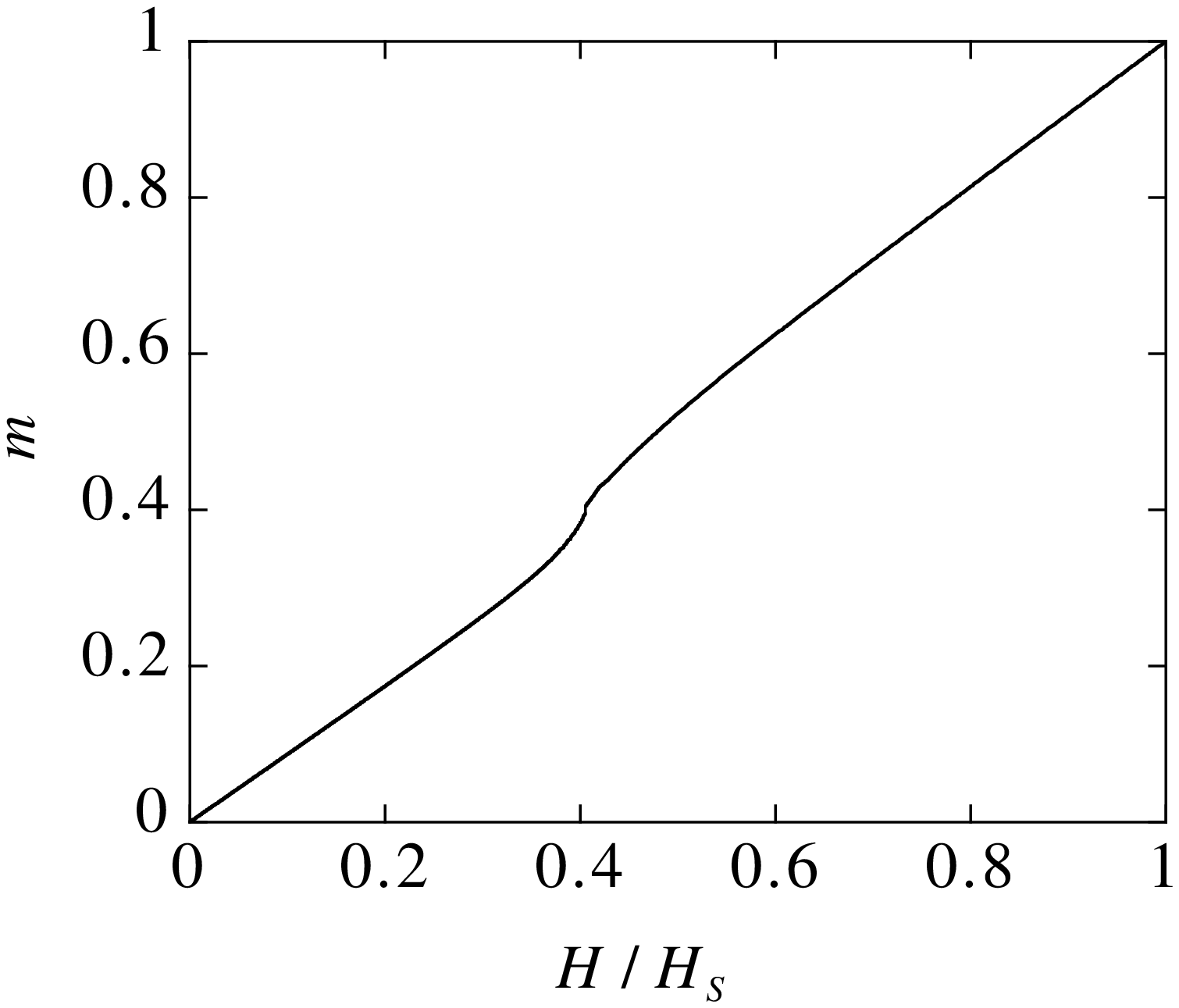}}
\begin{caption}
{The field dependence of the reduced magnetization $m$.}
\label{m_plot}
\end{caption}
\end{figure}

Figure~\ref{m_plot}, our main result, shows the field dependence of the 
reduced magnetization
\begin{equation}
m={1 \over L}\sum_{l=1}^{L}\cos \phi_l.
\end{equation}
A discontinuity ($\Delta m\approx 9\times 10^{-3}$) occurs at $H=0.405H_S$; 
this field is however somewhat larger than the $H=0.32 H_S$ where a 
magnetization jump is found experimentally~\cite{tanaka}, and there 
is a lot of fine structure not reported in Ref.\cite{tanaka}. 

We now describe the phases in more detail.

\subsection{Low-field region: $0<H<0.405H_S$}

The structures are helical, with three equivalent sublattices; 
each sublattice phase winds through $2\pi$ a total of $N/3$ times, where 
the total winding number $N$ is a multiple of 3, like the chain length $L$. 

\begin{figure}
\epsfxsize=110mm
\centerline{\epsfbox{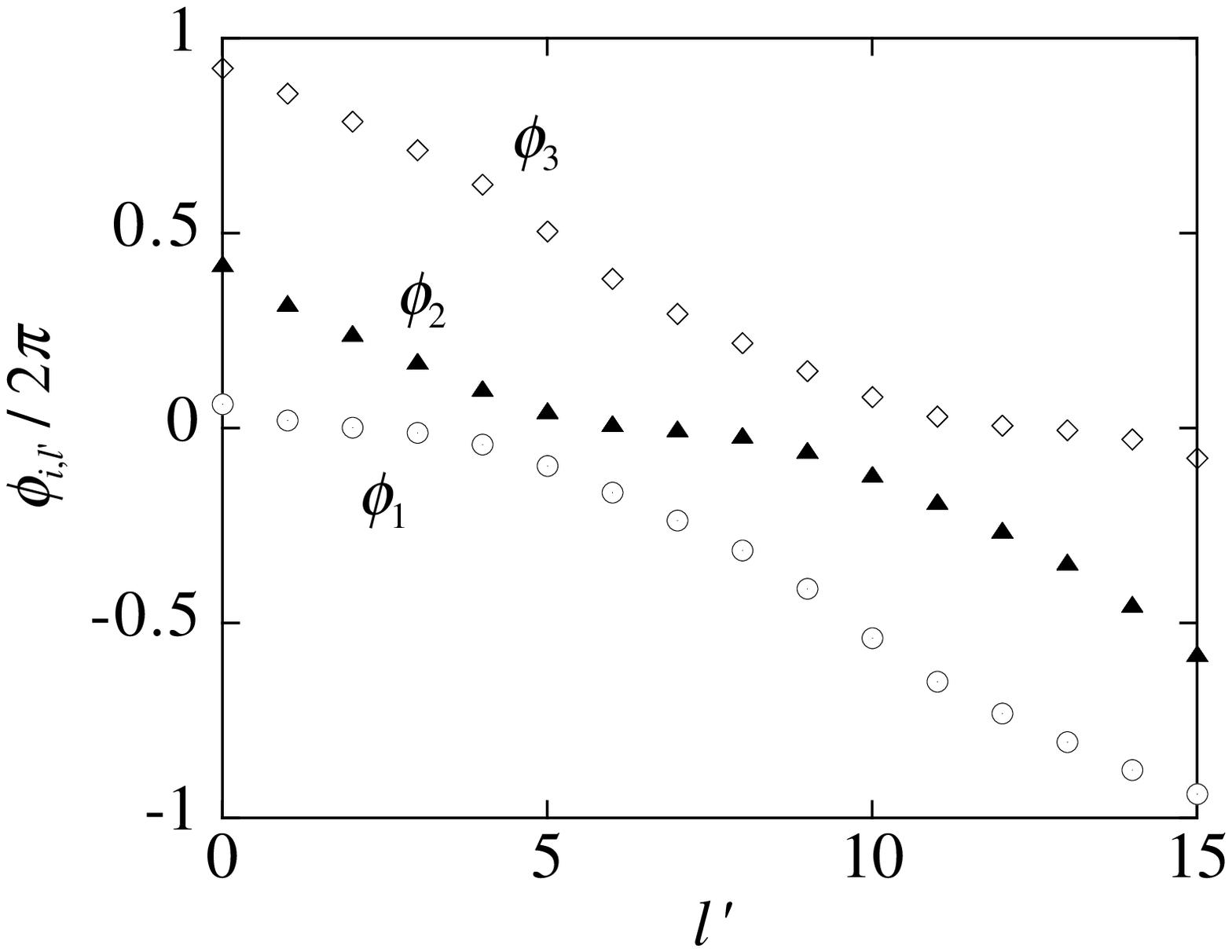}}
\begin{caption}
{The order parameter (the angles $\phi_{i,l'}$ in the three-sublattice
representation, as defined below Eq.~(4)) in the helical phase at 
$H=0.33H_S$.}
\label{IC1_plot}
\end{caption}
\end{figure}

The helical phase can be treated analytically at small $H$, as in 
Ref. \cite{nagamiya}. 
The magnetic field distorts the zero-field spin structure $\phi_l=q_0l$, 
modulating the phases.
In weak fields, 
\begin{equation}
\phi_l \approx ql+a \sin ql,
\end{equation}
where the optimal amplitude $a$ and wave number $q$ are field-dependent; 
they are determined so as to minimize the energy ${\cal E}$. 
As the magnetic field is increased, the spin structure distorts further as 
higher harmonics are generated. 
Figure~\ref{IC1_plot} shows the order parameters (the angles $\phi_{i,l'}$ 
in the three-sublattice representation) found from numerical solution of 
Eq.~(\ref{Euler}) at $H=0.33H_S$. 

\begin{figure}
\epsfxsize=110mm
\centerline{\epsfbox{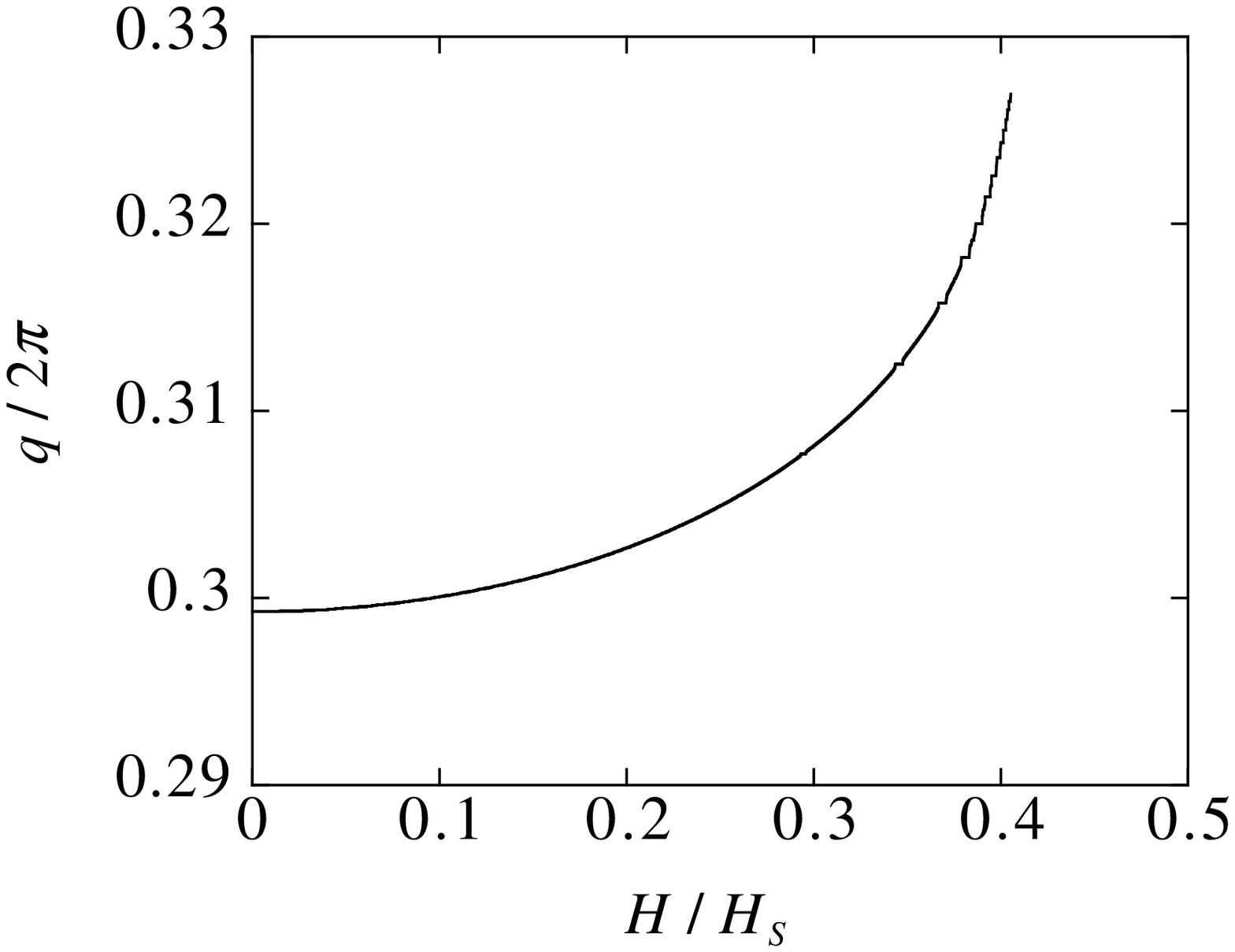}}
\begin{caption}
{The field dependence of the wave number $q$ in the low-field region
$0<H<0.405H_S$.}
\label{qlow_plot}
\end{caption}
\end{figure}

 The wave  number $q$ of the helical phase is 
\begin{equation}
q={ 2\pi\over3} \left( 1-{ N\over L}\right),
\end{equation}
in terms of the total winding number $N$ and the chain length $L$; 
one must distinguish between $q$ and the wave number $2\pi N/L$ of the chain.
>From Figure~\ref{qlow_plot}, at low fields $q$ increases with the field, 
initially quadratically, as the higher harmonics grow. 
At $H=0.293H_S$, however, a lock-in transition takes place to a helical C phase 
with $L/N=13$; 
this phase extends to $H=0.296H_S$ where the helical IC phase becomes optimal 
again. 
Other lock-ins occupying comparable field intervals occur at $L/N=16$, 19, 
{\it etc} in steps of 3, up to $L/N=52$ at the transition to the IC2 phase. 
Minor lock-ins (occupying much smaller field ranges) occur at $L/N=$ 29/2, 
35/2, 18, 20, 41/2, 21, 85/4, 91/4, 23, 47/2, 24, 53/2, 59/2 and 65/2; 
others likely exist but require a finer field step to be detected. 
Windows of the helical IC phase exist at higher fields, but they eventually 
become too narrow to detect (at our field step) and we see a staircase of C 
states which terminates at $H=0.405H_S$ in a strong first-order transition 
($\Delta m\approx 9\times 10^{-3}$) to the IC2 phase. 

\subsection{IC2 phase}
\begin{figure}
\epsfxsize=110mm
\centerline{\epsfbox{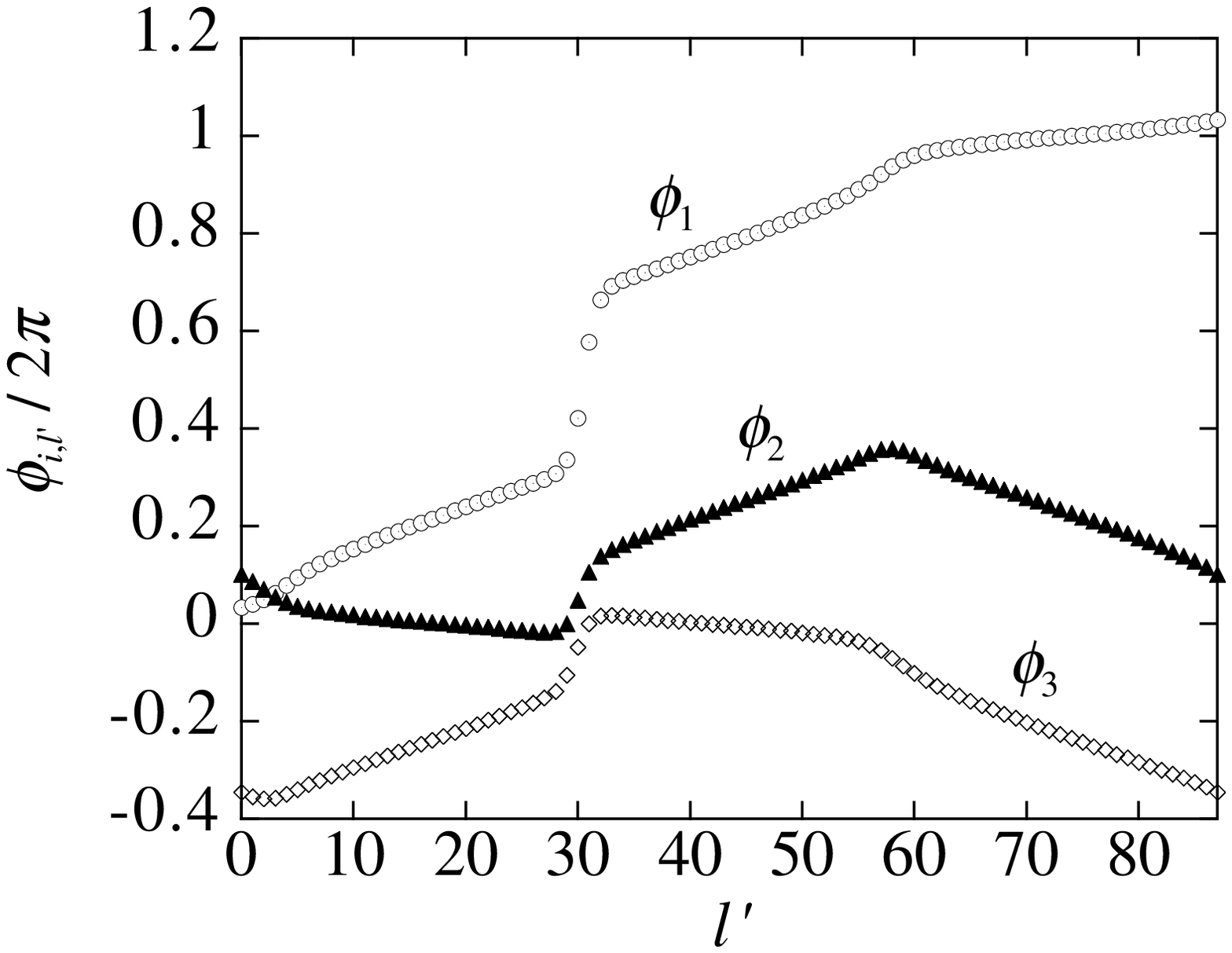}}
\begin{caption}
{The angles $\phi_{i,l'}$ in the IC2 phase at $H=0.41H_S$.}
\label{IC2_plot}
\end{caption}
\end{figure}

This phase is optimal in the very narrow range $0.405<H/H_S<0.419$. 
Figure~\ref{IC2_plot} shows the order parameters at $H=0.410H_S$. 
The structure resembles that of the helical phase (as shown in 
Figure~\ref{IC1_plot}); 
the three sublattices are however no longer equivalent, for the spins on 
only one of them wind through $2\pi$ (rather than all three as in the 
helical phase). 
The optimal chain period increases monotonically with the field, from 
$\approx 180$ to $\approx 1800$ at the transition to the IC3 phase; 
the latter transition is weakly first-order, with 
$\Delta m \approx 8\times 10^{-4}$. 

\subsection{IC3 phase}
\begin{figure}
\epsfxsize=110mm
\centerline{\epsfbox{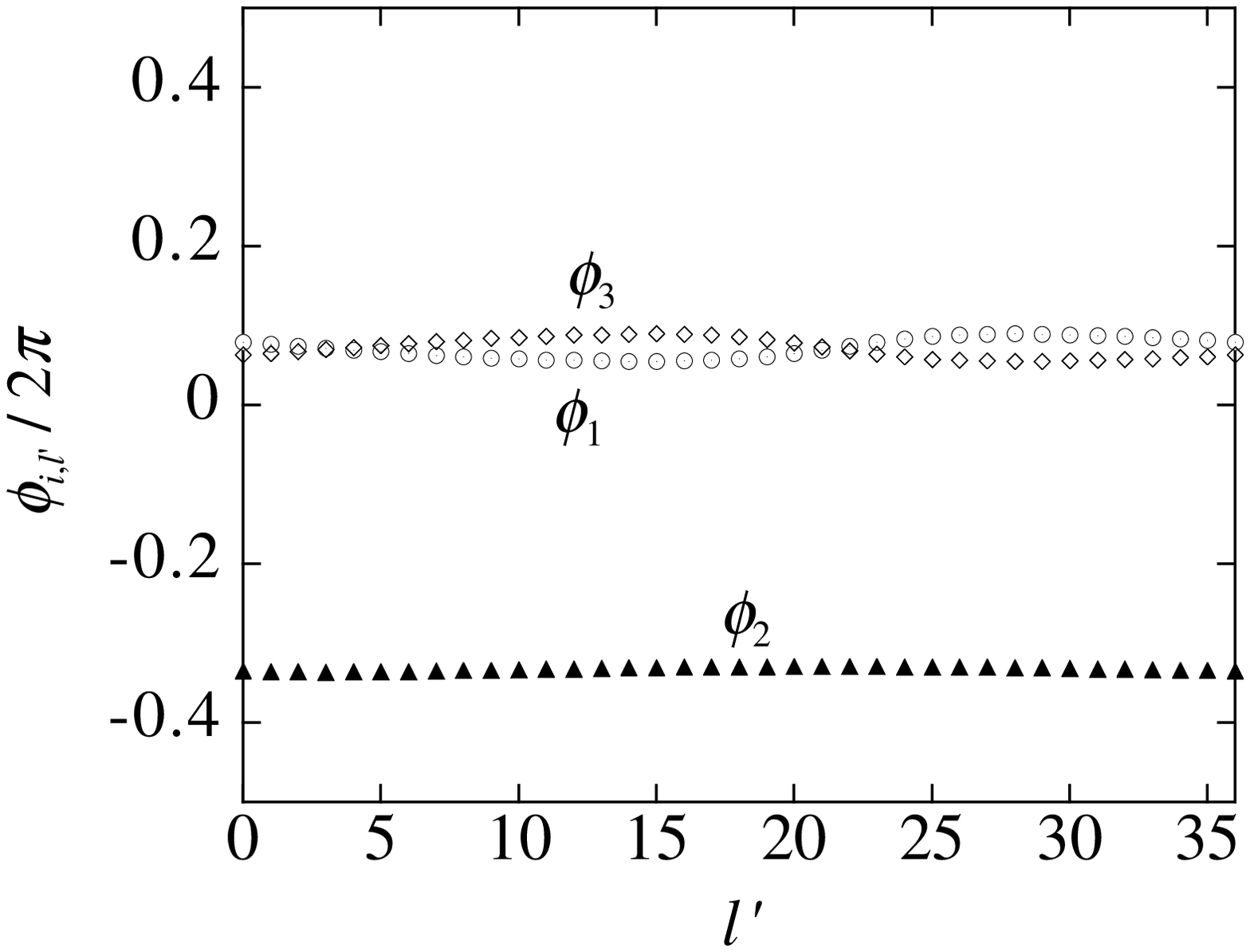}}
\begin{caption}
{The angles $\phi_{i,l'}$ in in the IC3 phase at $H=0.424H_S$.}
\label{IC3_plot}
\end{caption}
\end{figure}

This phase is also optimal over only a very small field range, 
$0.419<H/H_S<0.428$. 
The three sublattices are inequivalent, as in the IC2 phase, 
but here none of the three winds through $2\pi$. 
Figure~\ref{IC3_plot} plots the order parameters at $H=0.424H_S$; 
one sees that the IC3 phase is a rippled commensurate phase, in which the 
order parameter oscillates about a particular TAFM state. 
The IC3 state can be expressed in terms of the amplitude $a_i$ and the 
phase $\delta_i$ of the oscillation on the $i$th sublattice as 
\begin{equation}
\phi_{i,l'}\approx\phi_i+a_i\sin\left[\frac{2\pi l'}{(L/3)}+\delta_i\right],
\end{equation}
where $\phi_1=\phi_3$; 
the amplitudes are zero for the TAFM state in question. 
As the field increases, the oscillations become smaller and the period $L$ 
increases (from $\approx 80$ to $\approx 260$); 
correspondingly, the TAFM state becomes more favourable, but before it 
becomes optimal there occurs a very weak first-order transition 
($\Delta m \approx 2\times 10^{-5}$) to the fan phase. 
The IC3 phase (like the IC2 phase) requires a full numerical solution of 
Eq.~(\ref{Euler}); 
it should be possible however to demonstrate analytically that the TAFM 
state becomes unstable to these perturbations at some field $>0.428H_S$ 
(the upper limit of the IC3 phase). 

We point out a remarkable connection with the theory of fluctuations in the 
TAFM \cite{kawamura,chubukov}. 
Both thermal and quantum fluctuations break the classical degeneracy in the 
same way; 
for $H>{1\over3}H_S$, two angles are identical in the optimal state, say 
$\phi_1=\phi_3$ \cite{kawamura,chubukov}.
Our point is that this same state serves as the basis for the IC3 state in 
RbCuCl$_3$; 
the lattice distortion induces the spins to fluctuate spatially, and the 
energy is minimized when the fluctuations occur about the state with two 
angles identical, just as for thermal and quantum fluctuations. 

\subsection{High-field region: $0.428H_S<H<H_S$}
\begin{figure}
\epsfxsize=110mm
\centerline{\epsfbox{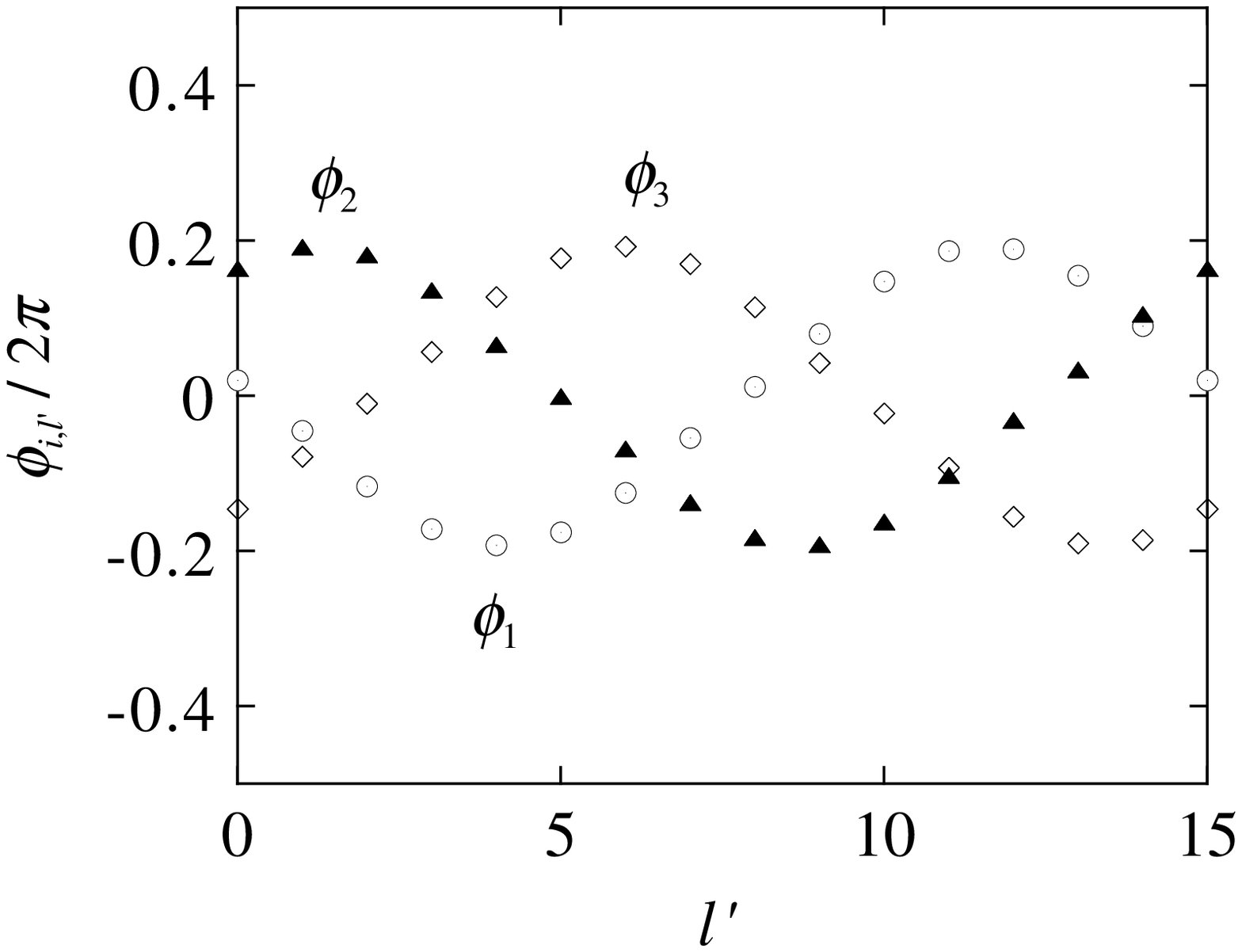}}
\begin{caption}
{The angles $\phi_{i,l'}$ in the fan phase at $H=0.66H_S$.}
\label{IC4_plot}
\end{caption}
\end{figure}

As shown in Figure~\ref{IC4_plot}, the spins oscillate about the field 
direction, without winding through $2\pi$; 
the three sublattices are again equivalent. 
The amplitude of the oscillation decreases as $H$ increases, going to zero 
at the transition, at $H_S$, to the aligned phase. 

\begin{figure}
\epsfxsize=110mm
\centerline{\epsfbox{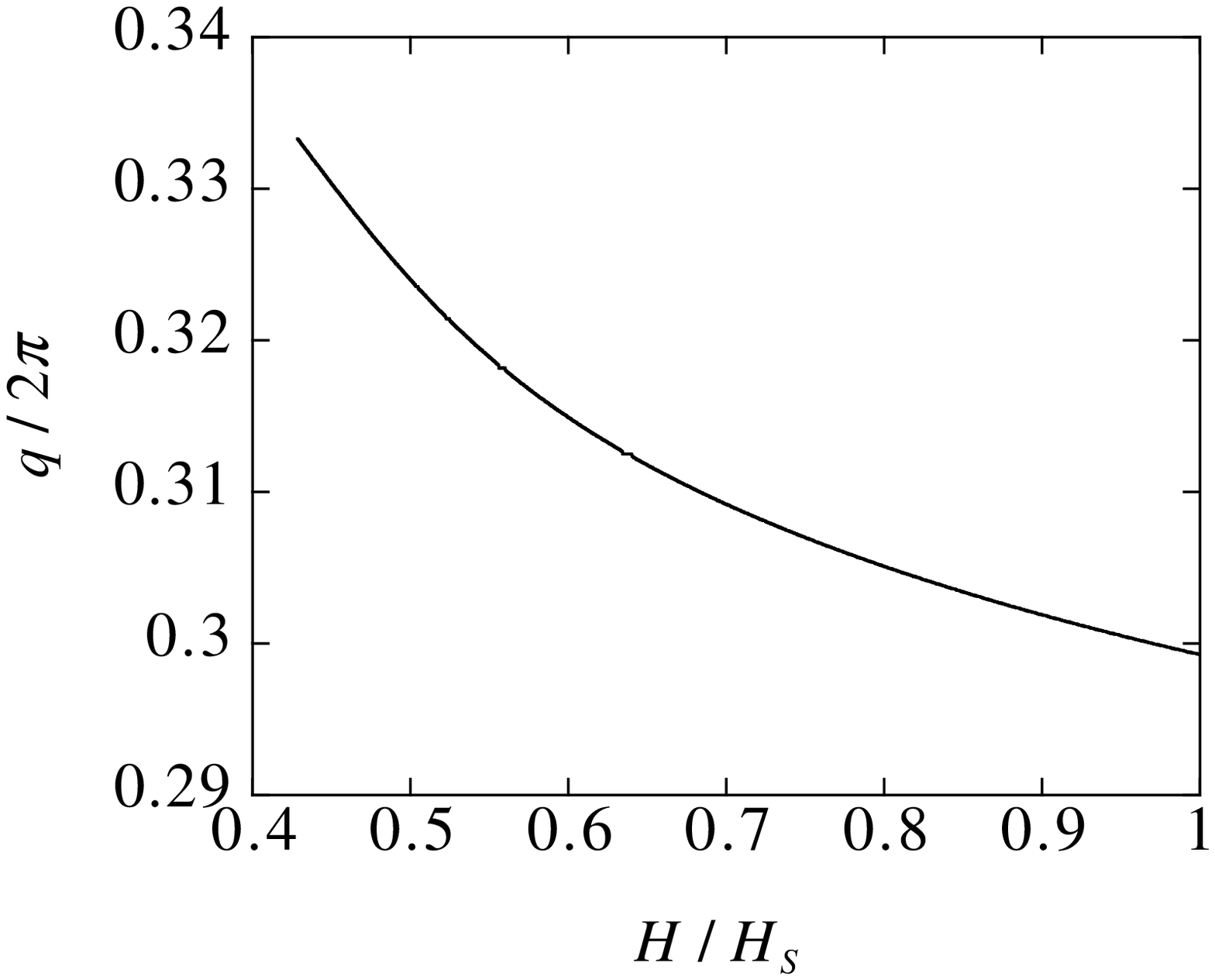}}
\begin{caption}
{The field dependence of the wave number $q$ in the high-field region 
$0.428H_S<H<H_S$.}
\label{qhigh_plot}
\end{caption}
\end{figure}

The wave number $q$ in this region is obtained from the chain period $L$ as 
\begin{equation} 
q=\left( {2\pi / 3} \right)\left( 1-3/L\right) . 
\end{equation}
As the field increases, the optimal chain length decreases from an 
estimated value $>10^5$ at the IC3-fan transition, to the interpolated 
value $\approx 29.4$ at $H_S$. 
Correspondingly $q$ decreases monotonically from $\approx 2 \pi/3$ to 
$\approx 1.88 (\approx 2\pi\times 0.299)$; 
Figure~\ref{qhigh_plot} gives numerical values. 
The structure is incommensurate in most of the fan region, but we find also 
narrow lock-in phases; 
these are less prominent than in the helical region at low fields. 
The widest C phase, with $L/3=16$ and width $0.006H_S$, occurs near 
$0.64 H_S$. 
Seven others occur at smaller $H$, at $L/3=22,$ 28 {\it etc}; 
smaller field steps would likely find many more. 

The transition at $H_S$ is second-order and so the fan phase can be treated 
analytically in this region, as in Ref.~\cite{nagamiya}. 
Expanding the energy function ${\cal E}$ in $\phi_l$, one finds  
\begin{eqnarray}
{\cal E} &=& S^2(4J_1+2J_1'-g\mu_{\rm B}H/S) \cr
&&+\frac{S^2}{2L}\sum_l 
\left\{\left[g\mu_{\rm B}H/S-\left(8J_1+4J'_1\right)\right]\phi_l^2
+8J_1\phi_l\phi_{l+1}+4J_1'\phi_l\phi_{l+2}\right\} 
+O(\phi^4). 
\label{E_2}
\end{eqnarray}
Introducing the Fourier transform
$\phi_l=\frac{1}{\sqrt{L}}\sum_q \phi_q e^{iql}$,
one can write (\ref{E_2}) as
\begin{equation}
{\cal E}  =  S^2\left(4J_1+2J_1'-g\mu_{\rm B}H/S\right)+
\frac{S^2}{2L}\sum_q\left\{g\mu_{\rm B}H/S-2[J(0)-J(q)]\right\}|\phi_q|^2 
+O(\phi^4) 
\label{E_q}
\end{equation}
where $J(q)\equiv 4J_1\cos q+2J'_1\cos 2q$.
Obviously the optimal state of (\ref{E_q}) is given by $\phi_q=0$ for all 
$q$ if $H>H_S=2[J(0)-J(q_0)]S/g\mu_{\rm B}$, where $q_0$ gives the minimum 
value of $J(q)$, {\it i.e.}, $\cos q_0=-J_1/2J_1'$.
Surprisingly, the optimal IC wave number in the limit $H\to H_S$ is identical 
to the value at $H=0$, even though the structures are quite different. 
Our analytical result 
\begin{equation}
H_S=(8J_1+ 2 J_1^2/J_1'+8J_1')S/g\mu_{\rm B} 
\label{HS}
\end{equation}
evaluates to $H_S=66$T on using the parameter values given below 
Eq.~(\ref{eq:H}). 
For $H\lesssim H_S$, the optimal solution is given by
\begin{equation}
\phi_l\simeq a \cos (q_0 l+\delta),
\end{equation}
where $\delta$ is an arbitrary phase; 
the optimal amplitude $a$ is determined by taking into account
the fourth-order terms in the energy function ${\cal E}$.

The difference between our result (\ref{HS}) and the field 
$(12J_1+6J_1')S/g\mu_{\rm B}$ (below which the aligned state is unstable 
with respect to TAFM states) is
$2\left(J_1-J_1'\right)^2 S /g \mu_{\rm B} J_1'$.
It is then obvious that the transition at $H_S$ with decreasing field is 
to a modulated state, except in the spatially isotropic case $J_1'=J_1$. 
The TAFM states are then rather fragile. 
They are unstable also at $H=0$ when $J_1'\neq J_1$, and we show above that 
they are never optimal for the anisotropy appropriate to RbCuCl$_3$; 
it remains to be determined whether they survive at intermediate fields for 
general anisotropy. 

By using the classical approximation, we have omitted the effects of quantum 
fluctuations. 
One can argue that quantum fluctuations are not important for RbCuCl$_3$, for 
their effect in CsCuCl$_3$ in transverse field is to give rise to a plateau 
in the IC wave number \cite{nikuni2}, rather than a new phase. 
On the other hand, quantum and thermal fluctuations in CsCuCl$_3$ stabilize 
states that are otherwise not optimal, in longitudinal \cite{nikuni} and 
transverse \cite{jacobs} fields respectively. 
We made a preliminary effort to understand the effects of quantum 
fluctuations, using a biquadratic term in the energy\cite{nikuni2}. 
Not unexpectedly, we find that the magnetization develops plateaus, in 
conflict with experiment, but we did not search extensively for other states. 

\section{Summary}
The classical theory of spin structures of RbCuCl$_3$ in a transverse 
magnetic field is surprisingly rich. 
We found four IC phases and as well two series of C phases. 
We showed that the wave number of the helical IC phase at $H=0$ is 
identical to that of the fan IC phase as $H\to H_S$, and we pointed out 
a remarkable connection between one of the IC phases and the effects of 
thermal and quantum fluctuations in the TAFM. 
We showed that the TAFM states of the isotropic case are fragile; 
infinitesimal anisotropy makes them non-optimal at both small and large 
fields, and they are non-optimal at all fields for RbCuCl$_3$. 
We found that the magnetization is discontinuous at $H\approx0.41H_S$, in 
qualitative agreement with experiment. 
Direct observation of the two intermediate IC phases and the C phases will 
be difficult; 
they occupy only small field intervals, and the large value of $H_S$ 
requires demanding pulsed experiments. 
Our predictions for the structure of the high-field fan phase are likely 
more easily tested. 

\bigskip
\begin{center}
{\bf Acknowledgements}
\end{center}
We thank H. Tanaka for useful correspondence.
This research was supported by NSERC of Canada and JSPS of Japan.

\end{document}